TITLE PAGE

# A versatile and robust microfluidic device for capillary-sized simple or multiple emulsions production

E. Teston*, V. Hingot, V. Faugeras, C. Errico, M. Bezagu, M. Tanter, O. Couture

Institut Langevin, Inserm U979, CNRS, ESPCI Paris, PSL Research University, 17 rue Moreau, 75012 Paris, France. * To whom correspondence should be addressed. E-mail: eliott.teston@chimie-paris.org, +33 6 10 96 28 44.

ORCID:
E. Teston: 0000-0002-6264-0607
V. Hingot: 0000-0003-2690-3171
V. Faugeras: 0000-0002-1343-1569
M. Tanter: 0000-0001-7739-8051
O. Couture: 0000-0002-8683-1424

**Abstract**

Ultrasound-vaporizable microdroplets can be exploited for targeted drug delivery. However, it requires customized microfluidic techniques able to produce monodisperse, capillary-sized and biocompatible multiple emulsions. Recent development of microfluidic devices led to the optimization of microdroplet production with high yields, low polydispersity and well-defined diameters. So far, only few were shown to be efficient for simple droplets or multiple emulsions production below 5 microns in diameter, which is required to prevent microembolism after intravenous injection. Here, we present a versatile microchip for both simple and multiple emulsion production. This parallelized system based on microchannel emulsification was designed to produce perfluorocarbon in water or water within perfluorocarbon in water emulsions with capillary sizes (<5 µm) and polydispersity index down to 5 % for *in vivo* applications such as spatiotemporally-triggered drug delivery using Ultrasound. We show that droplet production at this scale is mainly controlled by interfacial tension forces, how capillary and viscosity ratios influence droplet characteristics and how different production regimes may take place. The better understanding of droplet formation and its relation to applied pressures is supported by observations with a high-speed camera. Compared to previous microchips, this device opens perspectives to produce injectable and biocompatible droplets with a reasonable yield in order to realize preclinical studies in mice.

**Acknowledegments**
This work was supported principally by Plan Cancer, project UICT. This work was also supported by LABEX WIFI (ANR-10-LABX-24) within the French Program "Investments for the Future" under reference ANR-10-IDEX-0001-02-PSL*. The authors thank Jean Marie Chassot (Institut Langevin) Baptiste Heiles and Charlotte Constans for their helpful advices.

TEXT

## Introduction

Drug-delivery is a vast domain of research and it allows to envision more efficient treatments with reduced side-effects. It often implies the intravenous injection of carriers designed to be activated specifically in disease targets, such as cancerous tumors. Targeted drug-delivery can be performed with ultrasound, which can be focused with a submillimetric precision deep inside the body and can trigger the release or the disruption of acoustically-sensitive agents (Couture et al. Review 2014). This concept has shown enormous potential for the delivery of drugs (Gao et al. 2004), DNA (Suzuki et al. 2007) or dyes (Fokong et al. 2012).

As micrometric bubbles couple very efficiently with ultrasound (MHz range) to produce various mechanical and thermal local effects, several drug-delivery agents designed around microbubbles or gas-precursor droplets were conceived. A promising *in vivo* application is acoustic droplet vaporization (ADV) where microdroplets follow a liquid-to-gas phase shift, leading to the formation of large bubbles (around 25 µm), transiently occluding local microvasculature (Lin & Pitt 2013; Sheeran & Dayton 2014) and locally delivering its content (Couture et al. 2011 & 2012; Bezagu et al. 2014). Recently, a new way of realizing ADV, called acoustic cluster therapy (Sontum et al. 2015; van Wamel et al. 2016 a & b) has emerged and proven its efficiency to enhance vasculature permeability and extravasation of co-injected therapeutics (van Wamel et al. 2016b; Park 2016).

As described by our group (Couture et al. 2011 & 2012; Bezagu et al. 2014) and others (Fabiilli et al. 2014), a particular subclass of ultrasound-vaporizable agents are composite microdroplets. They are formed by a nanoemulsion of water in perfluocarbon (PFC), itself encapsulated in micron-sized droplets. The PFC forming the matrix of droplets can be vaporized by a microsecond-duration ultrasound pulse, hence destabilizing the droplet and allowing the inner and outer water phase to mix. We chose perfluorohexane (PFH) for its phase transition properties: as it vaporizes at 58°C, it allows us to produce stable emulsions at room temperature while its vaporization remains possible using clinical ultrasound devices. Moreover, these droplets also have the advantage to carry large payloads in each droplet and isolate the inner phase from the exterior through a hydrophobic and lipophobic PFC matrix. However, these composite droplets need to be sufficiently large, at least a few microns in diameter, to contain efficiently the nanoemulsion.

Conversely, before being activated by ultrasound, these composite droplets, like any intravenous drug-delivery carrier, need to be circulating. For agents injected intravenously, the upper size limit is that of the capillary diameter in order to prevent embolization of the microvasculature. A study of Wiedeman et al (1963) summarizes the diameter of capillaries in numerous species. In healthy human, normal capillaries have diameters ranging from 5 to 12 µm (Landau & Davis 1957). Earlier works showed that capillaries are 3.0 to 10.0 µm wide (Zweifach 1937) or that arterioles, capillaries and venous capillaries are about 5.0 or 6.0 µm in mice (Algire 1954).

Hence, in our view, having microdroplets of less than 5 µm is necessary for in vivo applications in order to efficiently reach peripheral circulation, avoid embolism and remain vaporizable by a clinical ultrasound scanner. Moreover, tightly monodispersed population of ultrasound-inducible particles were shown to be favorable as a therapeutic tool (Choi et al. 2010) and also make the therapeutic effect more predictable (Hingot et al. 2016). Unfortunately, producing uniform composite emulsions below 5 microns in diameter and in sufficient quantities for injection remains extremely challenging.

In previous works, microdroplets were prepared using microfluidic device in polydimethylsiloxane (PDMS), mixing flow-focusing and step emulsification as proposed by Cohen et al (2014). We also developed similar parallelized systems to enhance production rate (Cohen et al. 2014). However, simple systems had a low yield of production (less than 2500 droplets per second, while *in vivo* applications require around $10_8$ injectable microdroplets per mice) and parallelized ones were too sensitive to pressure conditions. This is especially critical as we need to inject a nano-emulsion inside the device if we desire multiple emulsions. This nano-emulsion of water in PFH has a mean hydrodynamic diameter of 230 nm and cannot be filtered before injection into a microfluidic device (Couture et al. 2012). Thus, any impurity or dust may clog some channels, leading to the unproductivity of the concerned system, an increase of pressure inside other channels and eventually a destabilization of the global production. Furthermore, PDMS has some drawbacks such as unstable wettability and a low robustness.

These issues could be solved through the numerous new developments in the field of microdroplet and microfluidic technologies. Progresses have been mostly driven by research in cosmetic or food science where production efficiency, monodispersity and stability remain the most important parameters (Zhu & Wang 2017; Lee et al. 2016;

Shang et al. 2017). Advances in these requirements were achieved through optimization of geometry of droplet production systems and adequate control of physico-chemical parameters such as wetting and shear rate, both for simple and multiple emulsion production (Zhu & Wang 2017; Lee et al. 2016; Shang et al. 2017). Some works aim at producing droplets for biological research (chemical synthesis and analysis, screening, enzyme kinetics, cell culture and sorting, protein crystallization (de Mello 2006; Janasek et al. 2006; Bezagu et al. 2014)) or drug encapsulation and delivery (Shang et al. 2017; Koster et al. 2008; Song et al. 2006). Many papers reported the efficient production of simple emulsions (up to 1.4 liters per hour) (Kobayashi et al. 2012) or monodisperse multiple emulsions (Kawakatsu et al. 2001; Utada et al. 2005), or even thermo-, chemo- or photo-triggered delivery with multiple emulsions (Lee et al. 2016). Yet, only few achieved the production of microdroplets compatibles with intravenous delivery (< 5 μm) with a low polydispersity as well as reasonable production rates.

In general, most microfluidics droplets production systems may be separated in three kinds: co-flow, cross-flow and flow-focusing geometries. These three methods allow the production of well-defined droplets depending on hydrodynamic parameters. However, the change of a parameter leads to the modification of droplets characteristics. Another technique as micropipetting solves this problem using gravity as an external force balancing interfacial tension (Christopher & Anna 2007). Yet, this technique is not suitable for micrometric dimensions as gravity becomes negligible compared to interfacial tension and viscous forces (Dangla et al. 2013).

Nakajima's team introduced many microfluidic systems allowing the production of droplets without the necessity of fine-tuning hydrodynamic parameters, and especially shear, by adapting microchip geometry and dimensions (Sugiura et al. 2002a; Kobayashi et al. 2005, van Dijke et al. 2010). These systems are based on the use of microchannels (MC) with defined dimensions coming out on large rivers with terraces (Fig. 1). The terrace drives the confinement of liquid between two horizontal walls. Once the liquid reaches the border of the terrace, it starts pulling itself outward, resulting in the formation of a spherical droplet with a lower superficial energy in the outer channel (Fig. 1). This method called MC emulsification has many advantages: the flow rate in MC and outer channel only have to be approximately defined; MC may be parallelized to obtain higher yields; geometry can be adapted to obtain desired droplet sizes and it requires a low energy input compared to other emulsification techniques (Sugiura et al. 2001). Indeed, droplet formation is the consequence of interfacial tension, the dominating force at this scale, which is itself consequent of the microchip geometry (Sugiura et al. 2001). Such spontaneous droplet formation devices enable the production of droplets with low dispersity, good yields (if there is parallelization), easy handling and robustness. Even if MC emulsification is less dependent on hydrodynamic parameters, some viscosity and flow rate considerations are needed in order to succeed in stable microdroplet production. Not only the design of the microchip is important, but it needs to be adapted to the characteristics of fluids that are used.

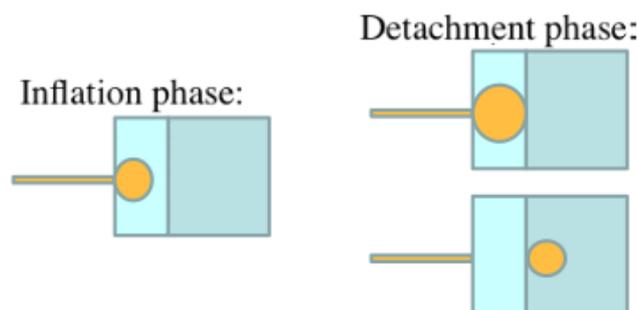

**Fig. 1** When escaping MC, the disperse phase follows first an inflation process when a pancake-like droplet is created on the terrace (left). In this confined space, the radius of droplet curvature is high, the droplet is in an unfavorable energy state as a spherical shape is desirable to lower interfacial tension. The pancake grows and eventually reaches the border of the plateau. There, interfacial tension is higher inside the terrace than outside. This energy imbalance will pull the droplet out into the deeper channel to transform into the more favorable spherical shape (right)

Terrace production happens if interfacial tension force is important enough compared to inertial and viscous forces. This mechanism was proposed by Sugiura et al. (2000) and suggest the importance of terrace dimensions on final droplet size. Indeed, the higher the plateau the lower the radius of droplet curvature inside the terrace and the harder and later the detachment of droplets. According to the literature (Sugiura et al. 2002a; Shui et al. 2011), droplet diameters have low dependency on terrace length but are proportional to terrace height. It was also shown that droplet diameter is proportional to the smallest slit dimension (in our case, terrace height) to a factor between 3.2 (Kobayashi et al. 2005) and 6.88 (Sugiura et al. 2002a) depending on the viscous ratios of the two phases (van

Dijke et al. 2010). Indeed, for a given terrace dimension, droplet size will be influenced by the viscosity ratio $\varepsilon$ of the two liquids (Equation 1). In this equation, the viscosities of disperse and continuous phases are denoted $\eta_d$ and $\eta_c$ (Pa.s), respectively.

$$\varepsilon = \frac{\eta_d}{\eta_c} \quad (1)$$

With a less viscous disperse phase, droplets have a larger diameter in the same continuous phase due to an increase of flow velocities at the exit of MC (Kawakatsu et al. 2001; Kobayashi et al. 2005). Van Dijke et al. (2010) proposed that a minimal viscous ratio is necessary to produce droplets and that a critical ratio exists above which droplet size is not anymore influenced by viscosity. Other studies proposed that this is the consequence of surfactant depletion around the future droplet: if the continuous phase is too viscous compared to disperse phase, droplets will form too rapidly compared to the necessary time for surfactant molecules to diffuse and adsorb on newly formed interfaces, destabilizing formed droplets (van Dijke et al. 2010; Vladisavljevic et al. 2011). These ratios seem device-dependent: their initial microfluidic system has a minimal viscous ratio of 0.48, but smaller terrace dimensions lead to a minimal ratio of 0.16. To explain this phenomenon, we propose that decreasing terrace dimension leads to a decrease of flow velocity at the exit of MC and an increase of interfacial tension force compared to inertial and viscous force, therefore permitting droplet production with a less viscous disperse phase.

All these studies convinced us of preferring MC microfluidic technologies that less rely on fine fluid flow tuning than what we previously developed. In this study, we introduce an appropriate microfluidic system for producing biocompatible PFH in water (PFH/W) or W/PFH/W ultrasound-vaporizable microdroplets. Our general aim is to develop new ultrasound-sensitive carriers to release active substances at a desired spot *in vivo*. We therefore developed a new design of parallelized microfluidic device taking advantage of MC and rupture confinement where the MC are fed and disposed through pressurized rivers (Couture et al. 2017; Martz et al. 2011). Glass lithography was preferred to PDMS in order to stabilize wetting properties and have robust and long lifetime microchips (Fabiilli et al. 2014). First, we present the production behavior of this device with a simple emulsion, before considering producing multiple emulsions. In both case we characterize obtained emulsions for given flow parameters and we try to understand phenomena that are responsible for changes in production regimes of this new device. Lastly, we need to ascertain that droplets vaporize easily using a clinical echograph. Thus, we ensured that we were able to realize ADV *in vitro* in the same conditions as with our previous microdroplets.

**Material and methods**

Chemicals:

*The water phase was made fluorescent by dissolving fluorescein sodium salt (Sigma–Aldrich Chemie GmbH, Germany) in mQ water at a final concentration of 1% (w/v). Glycerol (Amresco, USA) has been used to tune the water phase viscosity from 1 to 2.74 mPa.s (30 wt%). Surfactants have been used to stabilize generated droplets. The surfactants used were PEG-diKrytox (3% w/v in PFH) (Ran Biotechnologies, USA) and Poloxamer P188 (3% w/v in mQ water) (Sigma–Aldrich Chemie GmbH, Germany). Ultrasound vaporizable property of microdroplets was achieved using perfluorohexane (Acros Organics, USA) as the water non-miscible phase. Microchips were manufactured by Micronit Micro Technologies B. V. (Enschede, The Netherlands).*

Microchip design:

While we previously used PDMS microchip (Couture et al. 2011; Couture et al. 2012), we favored for this study glass lithography to have robust and long lifetime devices, as well as a good reproducibility and steady wetting properties. Nanochannels and MC were etched into a borosilicate glass wafer. The connection holes were drilled in a second borosilicate glass wafer using powder blasting techniques. Subsequently, these two wafers were aligned and thermally bonded together. The proposed microchips consist of two millimeter-deep rivers linked by 225 MC. MC parallelization has many advantages for the production of microdroplets: not only it allows production increase, but the clogging of one or few MC has no significant influence on the production of others. This is critical for producing W/PFH/W microdroplets as we inject a nano-emulsion of water in PFH, which we cannot formerly filter, inside these MC (Fig. 2). This potential clogging also justifies the use of the two parallel rivers, which remove dust particles before they are injected in the MC. A microfluidic flow control system (MFCS-EZ, Fluigent, France) was used to drive the liquid flows through the microchip. This set-up is hence entirely confined and the only communications between fluids and atmosphere are at the entrance of the flow control system and at the continuous phase outlet.

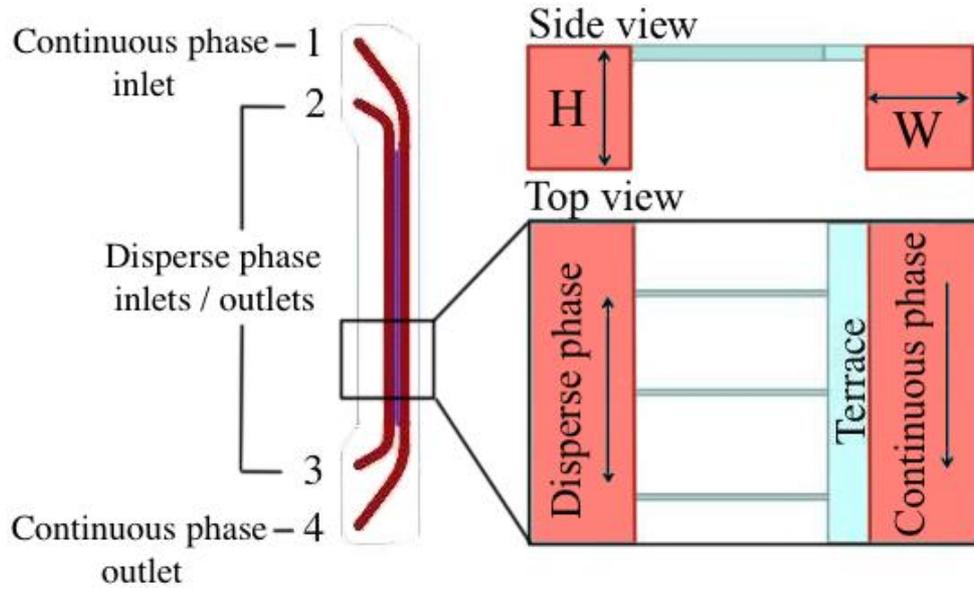

**Fig. 2** Schematic representation of our microfluidic device. Left: representation of millimeter-wide channels (red) and MC in between (blue). Continuous phase is flowing from inlet number 1 to outlet number 4 and disperse phase alternatively from 2 to 3 or 3 to 2. Pressures are controlled so disperse phase also goes through MC. Right: Side and top views of the production system. Schemes are not at scale: H = 0.1 mm and W = 1 mm, while MC and terrace height are 0.6 and 0.8 μm, respectively

The system is not symmetric: one channel (considered as the 'inner' channel) is made to receive the disperse phase, while the other one (the 'outer' channel), where the terrace is, is made to receive the continuous phase (Fig. 2).

In the proposed device, the two rivers are pressurized on both the inlet and the outlet. The flow in each river is determined by the pressure gradient between the inlet and outlet. Conversely, the flow inside MC between the two rivers is determined by the pressure gradient between the two inlets. Such system, with 4 pressurized containers allow the production to be entirely confined, except for the flow of air from the 4-channel pressurization system. Moreover, it allows a full automatization since the flow of the rivers can be inverted and the production can be performed in both directions. Applied pressure at the inlet number 1 corresponds to the external phase and will therefore be named ''external pressure'' while applied pressures at disperse phase inlets/outlets will be later named ''internal pressures''.

Considering geometry of previously described MC emulsification systems, we chose 250 μm long, 4.2 μm wide and 0.6 μm high MC and a slightly higher terrace (0.8 μm). Long MC are in favor of a more stable production of droplets (van Dijke et al. 2010; Sugiura et al. 2002b) and terrace height is linked to the desired droplet radius. Schematically, we can evaluate the final droplet volume by the terrace dimensions: as the liquid spread on the terrace in a disc-like shape, we chose dimensions so the volume contained in such a disc is the same as contained in a 4.0 μm diameter droplet (Equation 2).

$$R_d = \sqrt[3]{\frac{3}{16}HL^2} \quad (2)$$

H and L represent terrace height and length, respectively. Using Sugiura et al. predictive equation linking droplet diameter to terrace dimensions (Sugiura et al. 2002a) we find a predicted diameter of 4.8 μm, which is comparable to our simple estimation. As we will develop later, we experimentally obtained droplets with diameters ranging from 3.2 to more than 9 μm, depending on experimental conditions (characteristics of disperse and continuous phases, applied pressures).

Droplet formation observations:

We visualized microdroplet formation process using a set-up constituted of an inverted microscope (DMIL LED, Leica Microsystems) with a high-speed camera (Phantom v12.1, Vision Research, Ametek) and observations are consistent with the inflation and detachment model described in the introduction (Fig. 3):

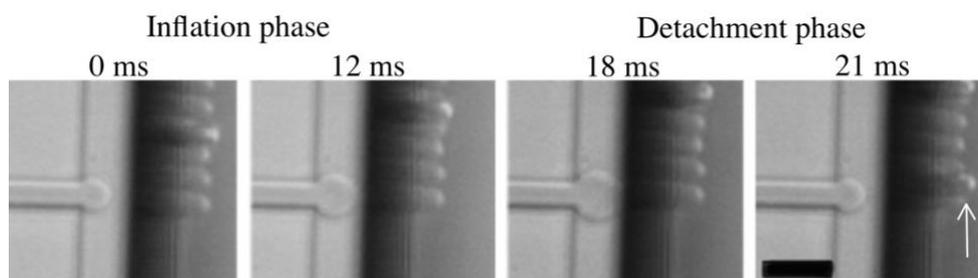

**Fig. 3** PFH microdroplet formation at the exit of a single MC (out of 225). Inflation phase: as PFH flows out of the MC on the terrace, a disc-like droplet is created. Detachment phase: once it reaches the border of the terrace, interfacial tension force pulls the droplet out of the terrace (white arrow), leaving a residual amount of disperse phase on the terrace (right image). Time at which photos were taken is written above each photo, pressure difference between internal and external channel is 50 kPa, scale bar represents 10 μm ($10^4$ fps)

Imaging this phenomenon gives the opportunity to measure the mean time between detachment of successive droplets in order to calculate mean flow velocity inside MC while changing pressures conditions. This will be further used to better characterize droplet production.

Emulsion production and characterization:

Production of simple PFH/W microdroplets was achieved by pushing pure PFH through MC, while using a 3% (w/v) P188 aqueous solution as the continuous phase. In order to obtain double emulsion droplets, our strategy was to first produce a nano-emulsion of water (containing 1% (w/v) fluorescein) in PFH before proceeding to a second emulsification using our micro-chip. First, the disperse phase was obtained using a Branson Sonifier 450 ultrasonic tip (Branson Ultrasonics, USA). Then, the disperse phase was pushed through the MC, as with the PFH in the first part of this study.
Droplet sizes and PDI were characterized by taking pictures after depositing 50 μL of a droplet solution in a p96 well. 4 photos of each well were taken with an inverted microscope (DMIL LED, Leica Microsystems) and size distributions were analyzed with a home written Matlab script (MathWorks, USA). At least three productions and analysis were realized for each condition.

Capillary number determination:

Determining mean diameters allows to calculate mean volume exiting MC per second and hence, disperse phase mean velocity and mean capillary number inside MC. Continuous phase velocity also depends on applied pressure. Calibration was realized measuring volume exiting the microchip for a given duration and applied pressure. Viscosities of PFH (6.70 mPa.s) and P188 water (1.66 mPa.s) and glycerol/water (1, 1.42, 1.84, 2.74 mPa.s for 0, 10, 20 and 30 % w/v, respectively) solutions were taken or calculated from the literature (Liu et al. 2011; Takamura et al. 2012) as well as interfacial tension of PFH against water (3.35 mN.m$^{-1}$) (Xiaonan et al. 2017).

Acoustic droplet vaporization:

A culture plate with $50.10^6$ microdroplets dispersed in 10 mL of glycerol in water (30% w/w, mimicking blood viscosity) is placed both at the focal distance of an acoustic transducer and a macroscope (x8). 2 μs acoustic pulses with various intensities were emitted while optic monitoring of droplet vaporization was realized.

**Results and discussions**

Simple emulsion production and size dependence with hydrodynamic or physicochemical parameters:

We first produced PFH/W microdroplets: we tested our microfluidic system with a large range of applied pressures both for inner and outer channels. It remains functional if inner channel pressures remained superior to outer ones and as long as the pressure difference is sufficient to observe droplet formation. Fig. 4 gives information about the dependence between droplet diameters, polydispersity index (PDI), applied pressures and hence flow velocities.

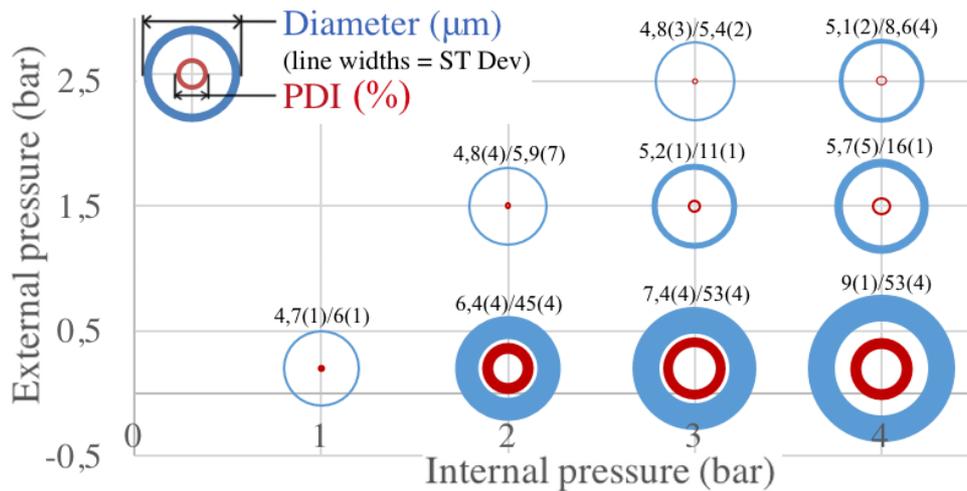

**Fig. 4** Diameters and PDI of PFH/W emulsions: characteristics of microdroplets produced with different applied pressures for simple emulsions of PFH in water. Diameters and PDI values are represented by blue and red circle diameters, respectively, while standard deviations are proportional to line widths (n≥3 for each condition). Diameters, PDI and standard deviations (written in brackets) are also displayed in number format on the top of each circle

On Fig. 4, we clearly see a trend between microdroplet characteristics and applied pressures. The higher ΔP, the bigger diameters and PDI. This is particularly obvious for an external pressure of 0.2 bar and internal pressures from 1 to 4 bars (lower circles on Fig. 4). Conversely, for a set internal pressure, increasing external pressure leads to smaller droplets and a lower PDI (see circles for 3 or 4 bars as internal pressures). Furthermore, we see that very close droplet diameters and PDI are obtained for different applied pressure. For instance, there is no significant difference between droplet characteristics produced at following internal/external pressure pairs (given in bars): 1/0.2, 2/1.5 and 3/2.5 or 3/1.5 and 4/2.5. Production rates range from 4 to $43.10^6$ droplets.min$^{-1}$ depending on applied pressures (Fig. 5).

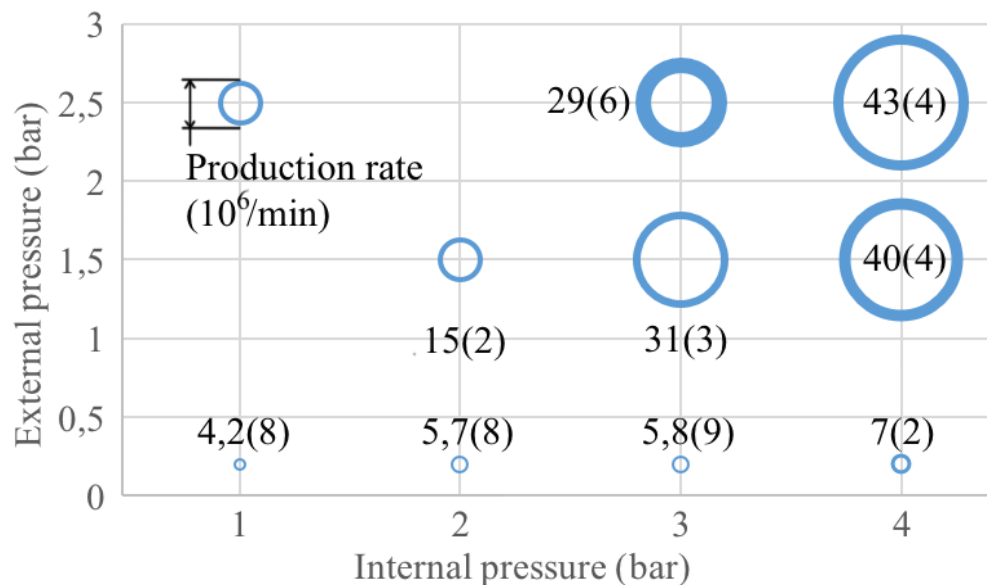

**Fig. 5** Production rate of PFH/W emulsions: amount of microdroplets produced with different applied pressures for simple emulsions of PFH in water. Production rate is proportional to circle diameters, while standard deviation is proportional to line width

For a given external pressure, increasing internal pressure leads to an increase of production rate, which is quite intuitive. However, one could think that increasing external pressure while keeping internal pressure constant would lower production rate, as pressure difference would decrease. Interestingly, for a given internal pressure, the production rate is at its lowest value when external pressure is low (0,2 bar). This is rather counter-intuitive but is explained by the fact that at these pressures, as droplet diameters is higher, much less droplets are produced for a volume of PFH exiting MC. Stolovicky et al. (2018) recently showed that for high rate of droplet production, they may accumulate where they form and interfere with next droplets to form and lead to an increase of diameter and PDI. In our case, this is more precisely due a coalescence phenomenon which will be discussed in the following paragraphs. A solution is to take away newly formed droplets, which is easy to realize with our design by applying a higher external pressure, inducing a higher continuous phase flow rate.

Fig. 6 illustrates the consequence of applying various pressures on obtained microdroplet characteristics. Three examples were chosen to visualize the evolution of microdroplet populations. Applied ΔP increases from left to right (0.5, 1.8 and 3.8 bars, respectively). In the first condition (ΔP = 0.5 bar, Fig. 6.a), the droplet population is homogeneous (PDI = 5.4 %) with the desired mean diameter. In the second condition (ΔP = 1.8 bar, Fig. 6.b), droplets still have a relatively low droplet diameter (< 10 μm) but became very heterogeneous (PDI = 45 %). With the highest ΔP (3.8 bars, Fig. 6.c), mean diameter gets closer to 10 μm, leading to unwanted sizes for *in vivo* applications.

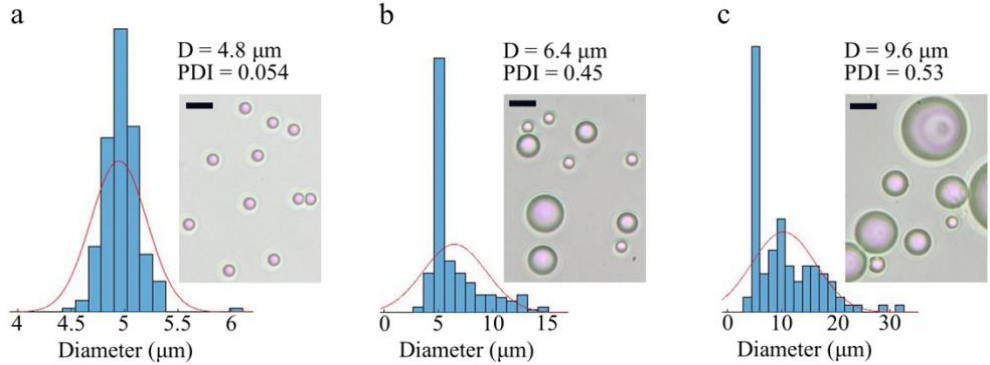

**Fig. 6** PFH microdroplet population diameters and PDI for various applied pressures. Internal/external pressures were set to 3/2.5 (a), 2/0.2 (b) and 4/0.2 (c) bars. Histograms of size distributions are represented next to representative pictures of microdroplets obtained at corresponding conditions. Scale bars represent 10 μm

Fluid flows in capillaries may be characterized using capillary numbers for disperse ($Ca_d$) and continuous ($Ca_c$) phases. Capillary number is the ratio between viscous and interfacial tension forces. Defining η (Pa.s) as the fluid viscosity, U (m.s$^{-1}$) as the mean fluid velocity and γ (N.m$^{-1}$) as the interfacial tension, capillary number is defined as follows:

$$Ca = \frac{\eta * U}{\gamma} \quad (3)$$

As both internal and external pressure seem to influence the characteristics of produced droplets, it is relevant to introduce a ratio comparing flow properties at the exit of MC. Capillary number ratio, as well as flow rate ratio, may be used for this purpose. Using previously introduced abbreviations, we can determine that capillary number ratio is proportional to velocity ratio:

$$\frac{Ca_d}{Ca_c} = \frac{\eta_d * U_d}{\eta_c * U_c} = \varepsilon * \frac{U_d}{U_c} \quad (4)$$

Fig. 7 shows that there is a linear relationship between droplet diameters and capillary number ratio. However, the correlation is low for PDI as we observe the apparition of a plateau (Fig. SI 1).

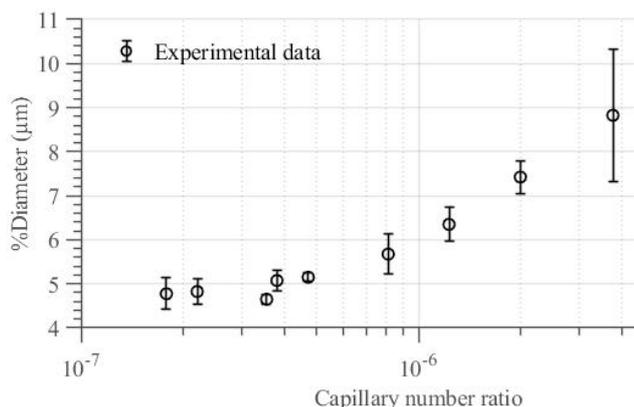
**Fig. 7** Mean PFH droplet diameter as a function of capillary numbers ratio

Many papers recently described systems that were relying on the geometry of the device to define droplet sizes (Sugiura et al. 2002a; van Dijke et al. 2010; Shui et al. 2011; Vladisavljevic et al. 2011). Yet, interestingly and similarly to our results, they obtain a high polydispersity and higher sizes when the inner phase flow rate increases (Shui et al. 2011; Vladisavljevic et al. 2011; Amstad et al. 2016). In a related manner to Fig. 7, they describe production regimes depending on the fluid velocity inside MC. Below a critical velocity $U_{cr}$, interfacial tension forces predominate and droplet diameter is constant (dripping regime) (Sugiura et al. 2002c), but above $U_{cr}$ the sum of viscous and buoyancy forces dominates over interfacial tension force and previous studies report a continuous flow of disperse phase, forming a big droplet at the MC exit (blow-up mode) (Vladisavljevic et al. 2011). Conversely to dripping mode, the droplet size become independent of the geometry of the microchip in blow-up regime. In our case, the critical velocity corresponds to a capillary ratio around $5\text{-}7.10^{-7}$. Furthermore, the production of microdroplets with a high polydispersity and higher sizes also happens when continuous phase viscosity increases and consequently continuous phase flow decreases (van Dijke et al. 2010). Thus, we assessed the influence of continuous phase viscosity on droplet characteristics. We tuned the viscosity by adding 0, 10, 20 or 30 % (w/v) of glycerol to the aqueous phase. For given applied pressures, the only visible consequence was a decrease of continuous phase velocity and, in turn, an increase of PDI and droplet mean diameters (Fig. SI 2). Indeed, if PDI and mean diameters are drawn against flow rate ratio (which is proportional to the capillary number ratio divided by the viscosity ratio), differences between each condition almost disappear (Fig. SI 3). This is consistent with the fact that flow velocity in outer channel is inversely related to fluid viscosity and it suggests that viscosity do not have a strong influence on droplet size if flow rate remains constant.

Interestingly, in our situation, we see that droplets that detach from the terrace are of similar dimension, and that increase in size is the consequence of coalescence of small droplets when they form too close to each other. *This is much different from previously described similar but larger production devices which suggest the straight formation of larger droplets at the exit of MC (van Dijke et al. 2010; Vladisavljevic et al. 2011).* Using a high-speed camera and decreasing the surfactant concentration by 100 times, we were able to unveil the underlying phenomenon responsible for the apparition of larger droplets (Fig. 8). This also explains why increasing external pressure, and thus continuous phase velocity, allows the production of smaller droplets: the increased flow in outer channels "flushes" formed droplets away, preventing them to remain where they appeared at the exit of MC. Additionally, increasing continuous phase velocity may help stabilizing newly formed interfaces by avoiding surfactant depletion (van Dijke et al. 2010; Vladisavljevic et al. 2011).

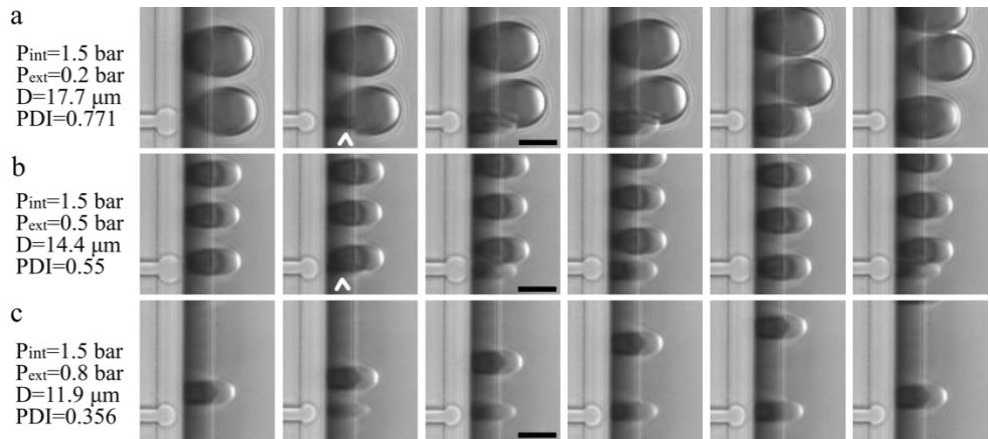

**Fig. 8** Production of PFH droplets in water with a low concentration of surfactant (0.03 % w/v) and various applied pressures. a) Photos were taken at 0, 0.3, 2.3, 6.5, 17.6 and 37.2 ms. b) Photos were taken at 0, 0.5, 3.2, 6, 14 and 19 ms. c) Photos were taken at 0, 6, 14, 22, 29 and 43 ms. Final mean diameters are 17.7 (a), 14.4 (b) and 11.9 (c) μm. a-b) White arrowheads show newly formed small droplets about to coalesce with following ones, leading to a new large one. Scale bars represent 10 μm

Fig. 8 depicts how external pressure (and hence continuous phase velocity) influences droplet size: a low external pressure such as 0.2 bar (Fig. 8.a) leads to larger droplets than higher external pressures and flows (Fig. 8.b and 8.c). The lower the flow, the lower the dragging force applied on droplets. Thus, a larger size is reached before droplet detached and flushes away from the exit of MC. Increasing external phase velocity drags smaller droplets faster, lessening their final diameter.

This observation is in favor of the statement that confinement rupture is a very stable and efficient technique to produce microdroplets with a specific diameter. Indeed, in our case, the phenomenon causing polydispersity increase is not related to droplet formation on the terrace but only to an insufficient speed of turnover of continuous phase outside the terrace (and also, in the previous experiment, to an insufficient concentration of surfactant). The reason why droplets coalesce may be that a high rate of droplet formation at the exit of MC leads to the formation of a new interface where surfactant has not enough time to diffuse. As a consequence, this new interface is not as stable as when production rate is low and coalescence is favored, as it has already been proposed (van Dijke et al. 2010; Wang et al. 2009). Nakajima's team suggested that, for their systems, continuous phase flow may only be useful for droplets recovery and do not influence droplet sizes in dripping regime (Vladisavljevic et al. 2011). However, for smaller systems as ours, it seems that droplet characteristics are proportional to capillary number ratio and hence related to both phase flows. This is converse with previously published results and may be related both to special chemicals we use and to the specificity of our microfluidic device.

Our initial aim was to develop a new microfluidic device allowing us to produce efficiently double emulsion for *in vivo* applications. Previously used parallelized systems, merging flow focusing and step emulsification (Cohen et al. 2014), were not optimized for this task. Indeed, the water in PFH nano-emulsion could clog some channels, leading to significant changes in flow parameters and hence in droplet characteristics. As this nano-emulsion cannot be filtered without altering its properties, we needed a system able to remain stable even with clogging of some channels. Furthermore, previous systems produced approximately 2500 droplets per second, which lead to more than 5 h of production for one mouse, if the production rate remained stable. *It was thus crucial for the development of preclinical assays to be able to produce intravenously injectable multiple emulsions in sufficient amount and reasonable duration*. As was demonstrated above, our microfluidic device can stand some pressure variations with no significant effect on droplet diameters as long as capillary number ratio do not change much. So, we realized a similar study relating pressure conditions to composite microdroplets characteristics.

Double emulsion production:

Similarly to simple emulsion production, mean diameters and PDI of multiple droplets were determined for many external and internal pressures (Fig. 9).

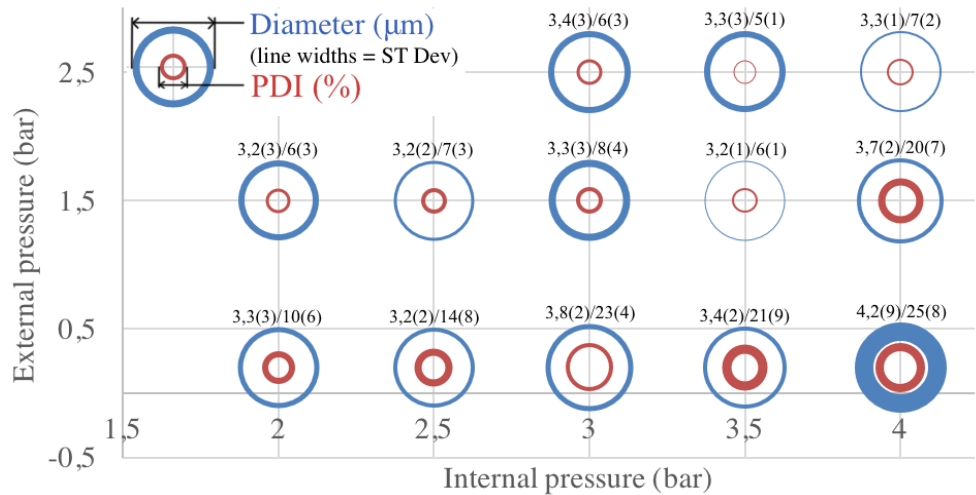

**Fig. 9** Diameters and PDI of W/PFH/W emulsions: characteristics of microdroplets produced with different applied pressures for double emulsions of water in PFH in water. Diameters and PDI values are represented by blue and red circles diameters, respectively, while standard deviations are proportional to line widths (n≥3 for each condition)

Conversely to PFH/W emulsion, we only observe a slight increase in droplet diameter for two conditions corresponding to the highest pressure ratios (4/0.2 and 4/1.5 bars for internal/external pressures). Hence, for W/PFH/W droplets, the dripping regime of this production system covers a larger range of pressures than for PFH/W emulsion. This may be explained by the fact that the nano-emulsion is more viscous than PFH, leading to reduced MC fluid velocities thus mimicking lower internal pressures. Still, high pressure ratios lead to significant increases of PDI. As for PFH/W emulsions, high internal pressures do not prevent from producing biocompatible droplets with diameters and PDI inferior to 5 μm and 10 %, respectively (for instance internal/external pressures of 4/2.5).

Production rates are not significantly different for all pressure conditions (Fig. SI 4) and range from more than 6000 to 25000 per second for droplets with PDI less than 0.08. This opens new experimental perspectives as we are now able to produce enough microdroplets in a half day to treat more than 5 mice, which used to necessitate at least two days of production. Furthermore, new microfluidic devices are easy to wash and may be reused for hundreds of productions with steady properties while previous PDMS microchip could not be used more than few times.

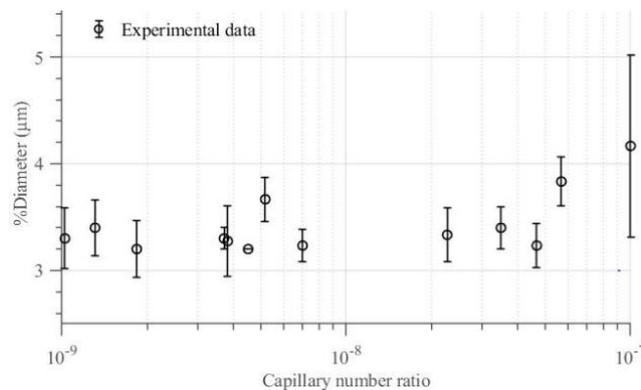

**Fig. 10** Mean W/PFH/W droplet diameter as a function of capillary numbers ratio

For multiple emulsion production, similar applied pressures to those used for simple emulsion production lead almost all to dripping regime. This is not surprising as the nano-emulsion is more viscous than pure PFH, decreasing flow velocity and hence capillary number inside MC. Our system did not allow us to apply higher pressures, and observe a larger blow-up regime domain, but diameters suggest it is beginning above capillary ratios of $5\text{-}6 \cdot 10^{-8}$ (Fig. 10), while PDI of multiple emulsions start to increase around $2 \cdot 10^{-8}$ (Fig. SI 5). We actually assessed different lengths of terrace and we eventually chose one which produces monodisperse and capillary sized multiple emulsions for our pressure range. Thus, this new microfluidic device has a dripping regime that is consistent with pressure we are able to apply (1-4 bars) and for described chemicals.

Fig. 11 illustrates the evolution of microdroplet populations for various applied pressures. ΔP increases from left to right (1.5, 2.5 and 3.8 bars, respectively). In the first condition (ΔP = 1.5 bar, Fig. 11.a), the droplet population is homogeneous (PDI = 5.6 %) with the desired mean diameter. In the second condition (ΔP = 2.5 bar, Fig. 11.b), droplets still have the desired droplet diameter (< 4 μm) but became more heterogeneous (PDI = 20 %). With the highest ΔP (3.8 bars, Fig. 10.c), mean diameter gets over 4 μm, and PDI increases to 25 %. Still, W/PFH/W have sizes and PDI that are much more compatible with *in vivo* applications than PFH/W emulsions, regardless of applied pressures.

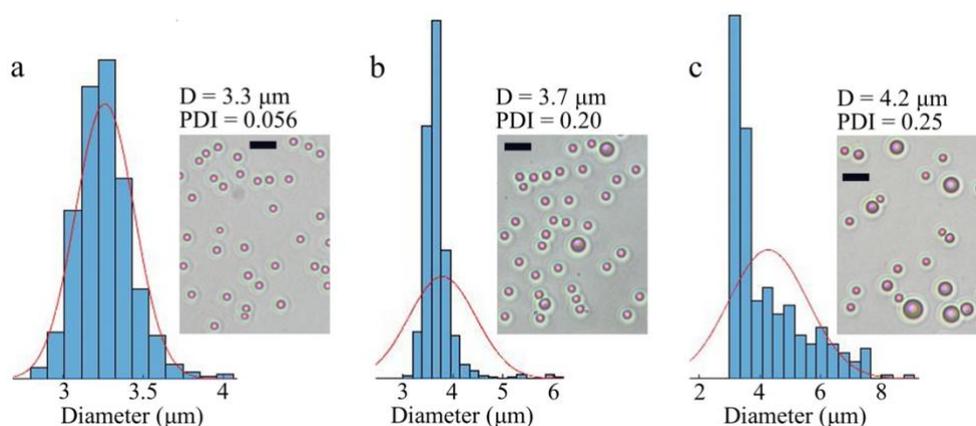

**Fig. 11** W/PFH/W population diameters and PDI for various applied pressures. Internal/external pressures were set to 4/2.5 (a), 4/1.5 (b) and 4/0.2 (c) bars. Histograms of size distributions are represented next to representative pictures of microdroplets obtained at corresponding conditions. Scale bars represent 10 μm

As demonstrated in the literature, the stability of dripping regimes of MC emulsification systems allows easy up-scaling of monodisperse microdroplet production (Kobayashi et al. 2012). In our situation, for both simple or multiple emulsion production, it is still necessary to control these parameters in defined ranges that are specific to production condition: microchip dimension and geometry, fluid properties and probably surfactants that are used. As we previously observed for PFH/W emulsions, high pressure ratios lead to the formation of larger droplets through coalescence phenomena. This is responsible for the increase of mean diameter and PDI and the coalescence phenomenon is illustrated in Fig. 12 for double emulsion production.

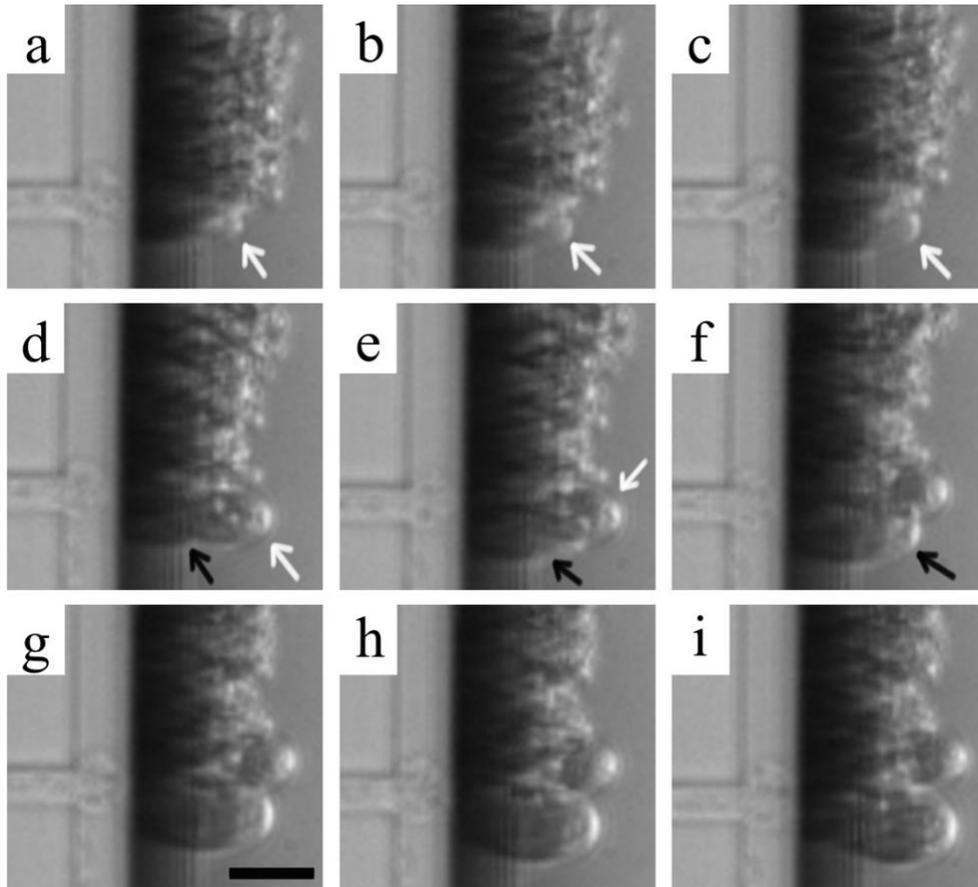

**Fig. 12** Observation of the formation of large microdroplets at the exit of MC with at 10$_4$ frames per second. Internal and external pressures were 3.5 and 0.2 bars, respectively. a-f) Formation of a first large droplet (white arrows) which eventually flows away (d) while a new large droplet starts to form (d, black arrows). Photos were taken at 0, 4, 8, 32.8, 38.5, 49.7, 56.3, 58.6 and 64 ms (from a to i, respectively). Scale bar represents 10 μm

Ultrasound vaporization:

Using a previously described set-up (Couture et al. 2011; Hingot et al. 2016) we observed *in vitro* droplet vaporization (Fig. 13). Fluorescence of fluorescein is quenched while fluorescein is confined at high concentration inside droplets (Nichols et al. 2012). When droplets vaporize, inner aqueous medium is spilled inside external water and fluorescein concentration drops, leading to a strong increase of fluorescence (fluorescence of fluorescein is approximately decreased 80 times at a concentration of 1%). This increase of fluorescence is seen after the ultrasound delivery pulse has been emitted (Fig. 13 middle and right). Even if we change our way of producing multiple microdroplets and used device, we kept acoustic vaporization properties of previously described emulsions (Hingot et al. 2016).

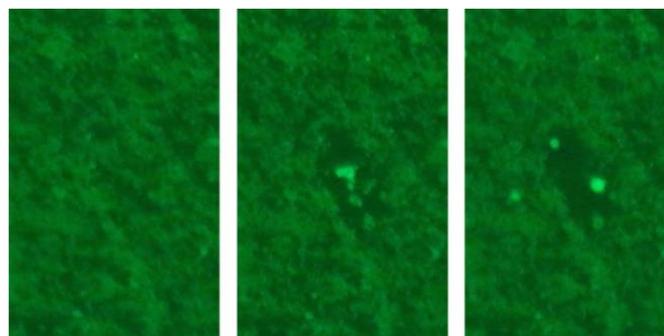

**Fig. 13** Optical monitoring of acoustic droplet vaporization. Images were taken before (left), 0.5 and 1 s (middle and right, respectively) after the focalized ultrasound push (2μs, 5 MHz, 3.1 MPa PnP) was emitted

**Conclusion**

An innovative design merging MC and step emulsification has been developed in order to overcome limitations we were facing when producing intravenous injectable multiple emulsions for further *in vivo* applications. This long lasting versatile microchip now allows us to produce enough microdroplets to consider intravenous drug delivery studies in small animal models. We also showed that this system may be used for both simple and multiple emulsions production, and that depending on chemicals that are used, and especially on their viscosities, droplet characteristics may be well tuned if applied pressures are within defined ranges. Observations with a high-speed camera were also realized in order to better understand microscale phenomenon governing droplet formation at a micron scale. Lastly, changing production technique and device did not altered acoustic vaporization properties of microdroplets, which remain inducible by a clinical ultrasound scanner.

**Conflict of interest**
MT, CE and OC hold a patent on a parallelized microfluidics droplet production device (PCT/FR2016052890).

**Supplementary**

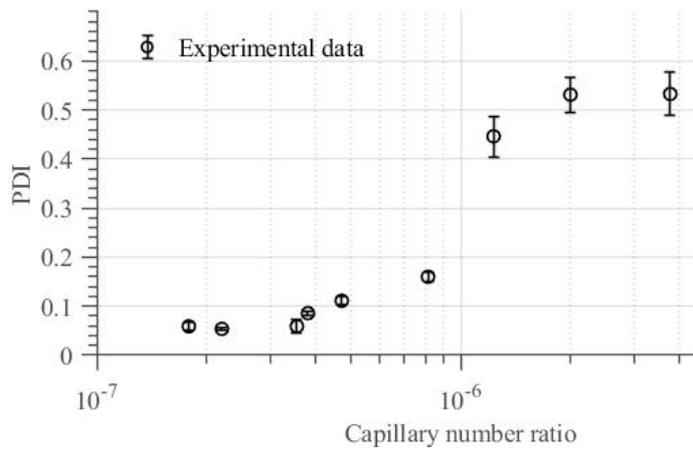

**Fig. SI 1** PFH droplet PDI as a function of capillary numbers ratio

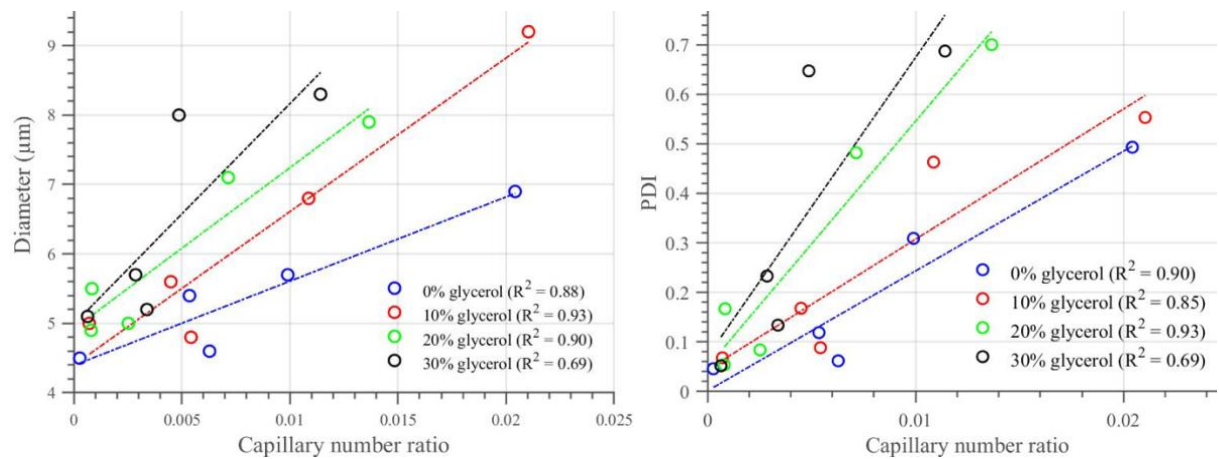

**Fig. SI 2** Droplet mean diameters (left) and PDI (right) as a function of capillary number ratio for increasing values of continuous phase viscosity

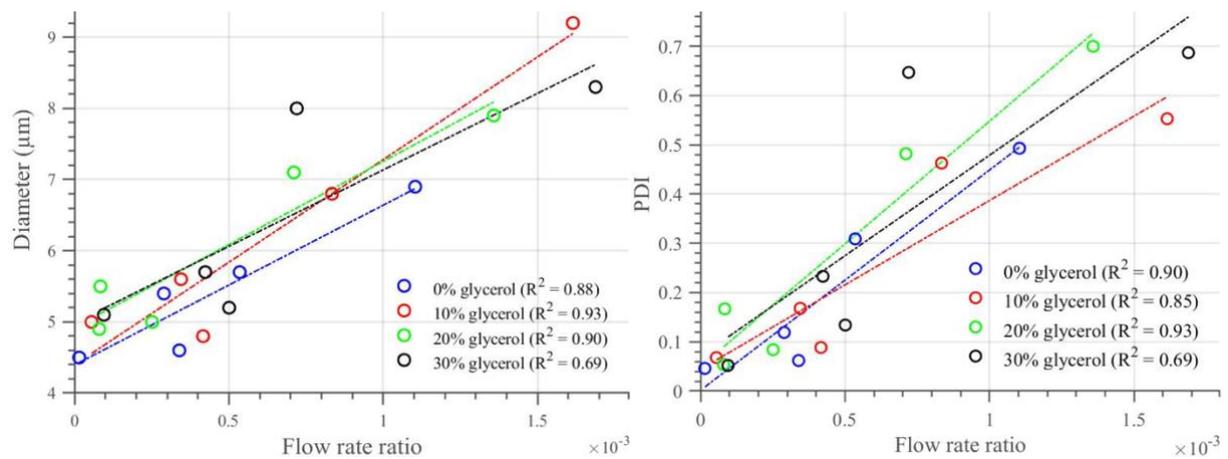

**Fig. SI 3** Droplet mean diameter (left) and PDI (right) as a function of flow rate ratio for increasing values of continuous phase viscosity

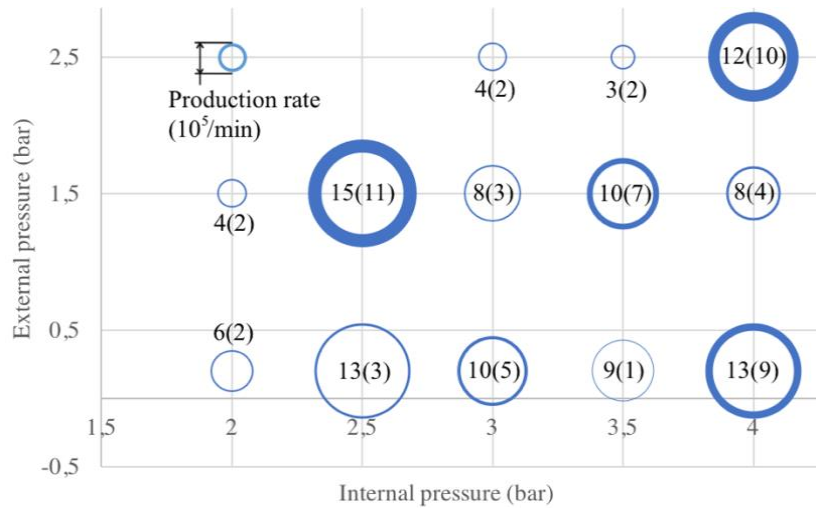

**Fig. SI 4** Production rate of multiple emulsions: amount of microdroplets produced with different applied pressures for emulsions of W/PFH/W. Production rate is proportional to circle diameters, while standard deviation is proportional to line width

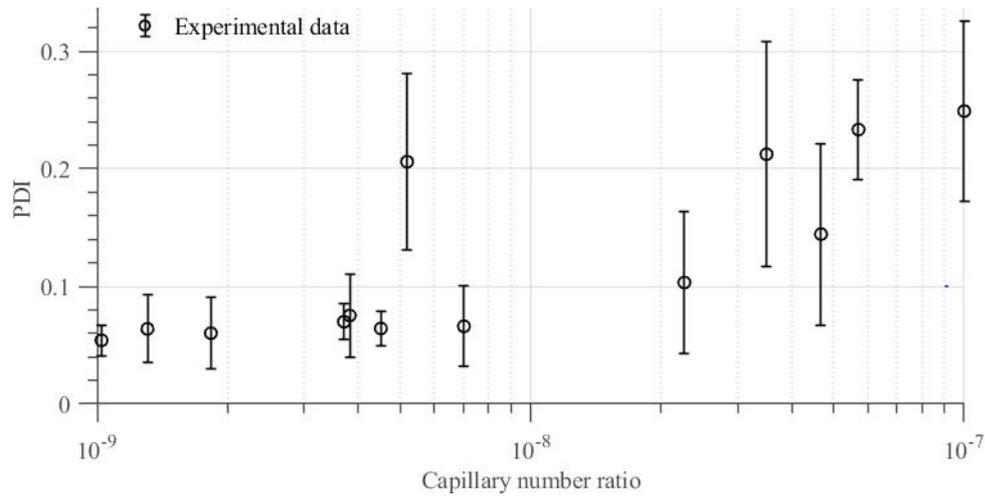

**Fig. SI 5** W/PFH/W droplet PDI as a function of capillary numbers ratio. Dripping regime takes place for capillary ratios inferior to $10^{-8}$ while blow-up regime takes place above this critical value